\documentclass[twocolumn,prl]{revtex4}
\usepackage{url}

\usepackage{amsthm}
\usepackage{amsfonts}
\usepackage{amssymb}
\usepackage{amsbsy}
\usepackage{graphicx}
\begin{document}

\title{Demonstration of Ultra Low Dissipation Optomechanical Resonators on a Chip}

\author{G.~Anetsberger, R.~Rivi\`ere, A.~Schliesser, O.~Arcizet \& T.~J.~Kippenberg}
\email{tjk@mpq.mpg.de}

\affiliation{
 Max-Planck-Institut f\"ur Quantenoptik, Hans-Kopfermann-Str.1, 85748 Garching
 }

\begin{abstract}
\noindent
Cavity-enhanced radiation-pressure coupling of optical and mechanical degrees of freedom gives rise to a range of optomechanical phenomena, in particular providing a route to the quantum regime of mesoscopic mechanical oscillators. A prime challenge in cavity optomechanics has however been to realize systems which simultaneously maximize optical finesse and mechanical quality. Here we demonstrate for the first time independent control over both mechanical and optical degree of freedom within \textit{one and the same} on-chip resonator. The first direct observation of mechanical normal mode coupling in a micromechanical system allows for a quantitative understanding of mechanical dissipation. Subsequent optimization of the resonator geometry enables intrinsic material loss limited mechanical $Q$-factors, rivalling the best values reported in the high MHz frequency range, while simultaneously preserving the resonators' ultra-high optical finesse. Besides manifesting a complete understanding of mechanical dissipation in microresonator based optomechanical systems, our results provide an ideal setting for cavity optomechanics.
\end{abstract}

\maketitle
\noindent
Over the past years it has become experimentally possible to study the coupling of optical and mechanical modes via cavity enhanced radiation pressure, which gives rise to a diverse set of long anticipated optomechanical phenomena\cite{BraginskyB77} such as radiation pressure driven oscillations\cite{KippenbergPRL05} and, as demonstrated in 2006, dynamic backaction cooling\cite{GiganNature06,ArcizetNature06,SchliesserPRL06,Dykman78,Braginsky02}. Moreover, this coupling can be exploited to perform highly sensitive measurements of displacement\cite{ArcizetPRL06, NJPsubmission} which may enable the observation of radiation pressure quantum backaction\cite{Caves81} or related phenomena. Major goals in the emerging field of cavity-optomechanics\cite{Kippenberg07,Kippenberg08}, such as ground-state cooling\cite{Wilson-RaePRL07,MarquardtPRL07} necessitate high optical finesse and high mechanical quality factors at mechanical oscillation frequencies exceeding the optical cavity's linewidth. While recently impressive progress has been made in creating experimental settings in which radiation pressure effects can be studied\cite{ArcizetNature06, GiganNature06, SchliesserPRL06, KlecknerNature06, CorbittPRL07,ThompsonA07}, a prime challenge still concerns attaining simultaneously high optical finesse and high mechanical $Q$-factors.
Nearly all approaches so far have combined traditional cm-sized optical elements (mirrors) with micro- or nanoscale mechanical oscillators which simultaneously act as mirrors\cite{ArcizetNature06, GiganNature06, KlecknerNature06, CorbittPRL07} (or as dispersive element\cite{ThompsonA07}). Other approaches have used the intrinsic mechanical modes of the optical elements\cite{Briant03}. Yet, in general, the required high reflectivity of the micro-element limits its dimension to wavelength size and thus sets an upper limit to the mechanical frequencies that can be achieved. Also, it is exceedingly difficult to attain high mechanical quality factors while maintaining high optical reflectivity as independent control of optical and mechanical degrees of freedom is generally not possible. Thus, although remarkably high mechanical $Q$-factors at low frequencies have been obtained\cite{ThompsonA07}, this and previous approaches\cite{ArcizetNature06, GiganNature06, KlecknerNature06, CorbittPRL07,SchliesserPRL06,Briant03} have so far not succeeded in combining mechanical $Q$-factors comparable to those achieved in the field of NEMS and MEMS\cite{Verbridge07, NaikNature06, ClarkIEEE05, EkinciRSI05, CraigheadScience00} with the best values of optical finesse\cite{KippenbergAPL04,RempeOL92,Grudinin06}.
But exactly the combination of both is important for applications such as low loss, narrow-band ``photonic clocks''\cite{KippenbergPRL05,RokhsariIEEE} and indispensable for fundamental studies aiming at approaching and detecting quantized motion in mesoscopic optomechanical systems\cite{SchwabPT05,Kippenberg07,Kippenberg08}.\\
\noindent
Here we show for the first time independent control over both optical and mechanical degree of 
\begin{figure*}[t!]
\includegraphics{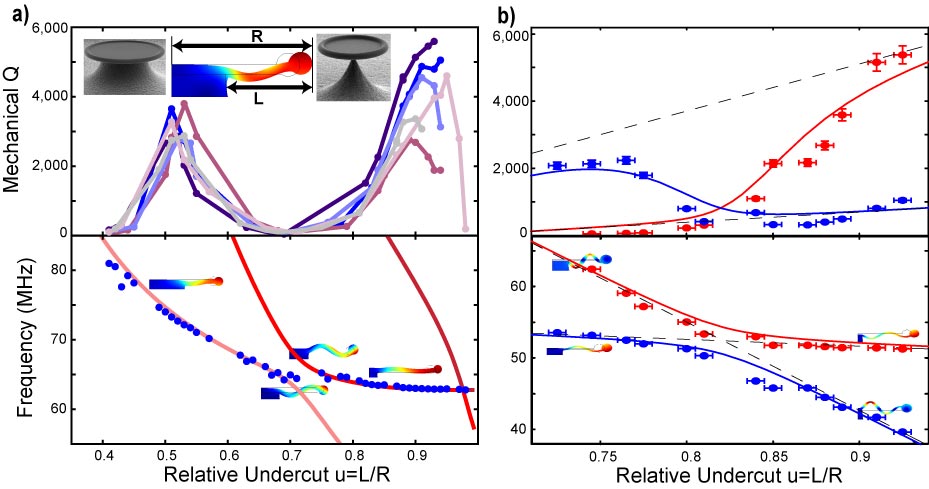}
\caption{\textbf{Observation of mechanical mode coupling. a} Mechanical $Q$-factors (upper panel) and frequencies (lower panel, where solid lines denote results of an FEM simulation) of the radial breathing mode (inset) for varying relative undercut $u=L/R$. The $Q$-factors were found to be remarkably reproducible for six different samples and strongly geometry dependent due to intermode coupling. \textbf{b} $Q$-factors (upper panel) and frequencies (lower panel) of radial breathing mode and a flexural mode of one toroid reveal an avoided crossing confirming that the dispersion lines of both modes do not cross. The mode patterns (radial and flexural modes) hybridize when approaching the coupling region while the corresponding mode patterns switch dispersion lines during the avoided crossing. A coupled harmonic oscillator model (solid lines: coupled $Q$-factors and frequencies; dashed lines: bare $Q$-factors and frequencies) allows an excellent fit to the data.}
\label{f:1}
\end{figure*}
freedom in \textit{one and the same} microscale optomechanical resonator. The demonstration of mechanical normal mode coupling\cite{Mindlin51} within a micromechanical device and the concomitant geometry dependence of clamping losses allows a \textit{quantitative} understanding of mechanical dissipation. We demonstrate monolithic spoke-supported silica resonators comprising a toroidal boundary which allows ultra-high optical finesse ($>10^6$)\cite{KippenbergAPL04} rivalling the best values obtained in Fabry-Perot cavities\cite{RempeOL92}. Independent control of their mechanical properties leads to strongly reduced clamping losses allowing for unprecedented mechanical quality factors (e.g. 80,000 at 38 MHz) -- rivalling the best published values of strained silicon nitride nanoresonators\cite{Verbridge07, NaikNature06} as well as radial contour-mode disk resonators\cite{ClarkIEEE05} at similarly high frequencies which allow entering the resolved sideband regime\cite{SchliesserNP08,Regal08} using state-of-the-art optical cavities. It is shown that the measured $Q$-factors are only limited by temperature dependent intrinsic dissipation\cite{Pohl02,Vacher05}, which can be reduced by low temperature operation. Moreover, the observed geometry dependent loss mediated by normal mode coupling may be of relevance across a wide range of micro- and nanomechanical oscillators, for which a detailed understanding of dissipation is lacking. 

\section{Results}
\noindent
Starting point of our analysis are toroidal silica microcavites\cite{ArmaniNature03} that intrinsically combine ultra-high-$Q$ optical whispering gallery modes with around 20 mechanical modes in the 0-100 MHz range\cite{NJPsubmission} which are observable in an
interferometric readout using the structure's optical modes (see methods). Recent work has shown that the intrinsic coupling of optical and mechanical modes of toroidal microcavities via radiation-pressure can give rise to the effect of dynamical backaction\cite{KippenbergPRL05}, which allows realization of narrow bandwidth photonic oscillators\cite{RokhsariIEEE} or photonic RF down-converters\cite{Hossein08} as well as radiation pressure cooling of a mechanical oscillator\cite{SchliesserPRL06}. The radial breathing mode (cf. Fig. 1a) in particular exhibits strong optomechanical coupling, low effective mass ($\approx\,10^{-11}\,\mathrm{kg}$) and high frequency enabling the first demonstration of resolved sideband cooling of a micromechanical oscillator\cite{SchliesserNP08}. To understand their mechanical $Q$-factors which have remained completely unexplored so far, we first present a study of mechanical dissipation in microtoroids and subsequently show how the results from this study can be harnessed to devise structures with unprecedentedly low dissipation in the frequency range above $30\, \mathrm{MHz}$.\\

\subsection{Observation of mechanical normal mode coupling}
In order to assess the contribution of clamping losses, the diameter of the silicon pillar holding the silica resonator (cf. Fig. 1a) is varied. To this end consecutive $\mathrm{XeF_2}$ etching cycles, undercutting the silica structure but leaving the silica itself unaffected, are applied. Thus, the relative undercut $u=L/R$ ($R$: radius of the cavity, $L$: length of the free standing membrane, cf. inset of Fig. 1a) can be controlled. The dependence of the measured $Q$-factors of the radial breathing mode (RBM) on the relative undercut is depicted in Fig. 1a. Intuitively, it may be expected that higher $Q$-factors are attained for larger undercut due to a reduced clamping area. Interestingly, however, the $Q$-factors of six different microresonators of similar size show a strongly non-monotonous dependence on the relative undercut which contradicts the simple expectation of smaller losses for smaller clamping areas. Moreover, the behaviour is remarkably reproducible for the different samples. A plot of the measured frequencies along with an FEM simulation (see methods) of the toroids' radially symmetric mechanical modes depicted in Fig. 1a shows excellent agreement of measured and simulated frequencies. Furthermore, the FEM simulation indicates an avoided crossing between different mechanical modes as first predicted by Mindlin in 1951\cite{Mindlin51}.
A measurement with a different sample allows \textit{directly} observing this avoided crossing by employing highly sensitive quantum limited optical displacement sensing (see methods) which allows to also monitor the flexural modes exhibiting weaker optomechanical coupling. The frequencies and the corresponding $Q$-factors of both the radial breathing mode and a flexural mode obtained in this second measurement are shown in Fig. \ref{f:1}b. The frequencies of both modes decrease for larger relative undercut, where the flexural mode shows a steeper slope. Thus, it can be observed that the eigenfrequencies of both modes approach each other but, as predicted by FEM simulations, they do not cross. Moreover, both mode patterns hybridize and eventually swap their dispersion lines during the avoided crossing (inset of Fig. \ref{f:1}b).
It is exactly in the region in which both modes couple, where the $Q$-factor of the RBM is found to be strongly reduced.

\noindent
This distinctive behaviour can be well explained with a generic model of two coupled harmonic oscillators $x_\mathrm{r/f}$, representing the bare radial (r) and flexural (f) mode, with angular frequencies $\Omega_\mathrm{r/f}$ and damping rates $\Gamma_\mathrm{r/f}$. An offset of the silica torus from the equatorial plane of the silica disk\cite{RokhsariIEEE} as well as the 
\begin{figure}
\includegraphics[scale=.74]{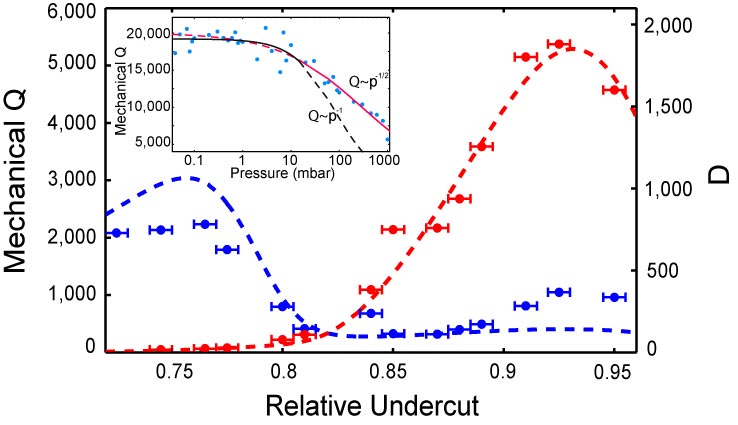}
\caption{\textbf{Linear relation between $D$ and measured $Q$-factors.} Mechanical $Q$-factors (dots with error bars, left axis) and corresponding simulated $D$ values (dashed lines, right axis) for the radial breathing and a flexural mode of an on-chip toroid show a linear relation in the parameter range depicted where $Q\approx 3 \cdot D$. Inset: Mechanical $Q$-factors depending on the background gas pressure. Below 1 mbar the gas damping contribution becomes negligible. The black and red lines correspond to fits in the molecular ($Q\propto p^{-1}$) and viscous ($Q\propto p^{-1/2}$) gas damping regime.}
\label{f:2}
\end{figure}
single-sided clamping of the disk to the silicon pillar gives rise to an appreciable coupling rate $g$ between both modes. The coupled eigenvalues of the system then read

\begin{eqnarray}
\lambda_\pm &=& \frac{\Omega_\mathrm{r}+\Omega_\mathrm{f}}{2} + i \frac{\Gamma_\mathrm{r} + \Gamma_\mathrm{f}}{4} \nonumber \\
&\hphantom{=}& \pm \sqrt{\left(\frac{\Omega_\mathrm{r} - \Omega_\mathrm{f}}{2} + i \frac{\Gamma_\mathrm{r} - \Gamma_\mathrm{f}}{4}\right)^2 + \frac{g^4}{4\Omega_\mathrm{r} \Omega_\mathrm{f}}}
\label{coupledequation}
\end{eqnarray}

\noindent
and thus the coupled eigenfrequencies $\Omega_\pm/2 \pi=\mathrm{Re}\left(\lambda_\pm\right)/2 \pi$  and $Q$-factors $Q_\pm=\mathrm{Re}\left(\lambda_\pm\right)/2\,\mathrm{Im}\left(\lambda_\pm\right)$ can be extracted. Assuming a linear dependence of the bare angular frequencies $\Omega_\mathrm{r/f} = \Omega_\mathrm{r/f}^{(0)}+\Omega_\mathrm{r/f}^{(1)} \, u$ and $Q$-factors $Q_\mathrm{r/f} = Q_\mathrm{r/f}^{(0)}+Q_\mathrm{r/f}^{(1)}\, u$ on the undercut $u$, these can be used to asymptotically fit the data (cf. Fig. 1b). Using the values obtained from this first fit, only $g$ remains as free parameter in equation (\ref{coupledequation}). A least square fit to both the measured frequencies and $Q$-factors yields a coupling rate of $g = 14\, \mathrm{MHz}/2\pi$, much larger than the damping rates $\Gamma_\mathrm{r/f}/2\pi$. The corresponding coupled frequencies $\Omega_\pm/2\pi$  and $Q$-factors $Q_\pm$ match the data remarkably well (cf. Fig. 1b) confirming that indeed both modes behave as two coupled harmonic oscillators giving rise to an avoided crossing between the RBM and the low-$Q$ flexural mode.
Moreover, the data presented here, to the authors' best knowledge, manifest the first direct observation of normal mode coupling of two different mechanical modes within a micromechanical resonator.\\

\subsection{Quantifying the clamping losses}
For the $Q$-factor of the RBM the avoided crossings are detrimental. The hybridization with the low-$Q$ flexural modes strongly reduces the mechanical quality of the RBM in the coupling region due to enhanced displacement amplitudes in the clamping area. Drawing on the quantitative power of FEM simulations, it is indeed possible to show that the displacement amplitudes at the pillar strongly influence the mechanical $Q$-factor. In order to study--and elucidate--the role of clamping losses the parameter $D$, defined as: 
\begin{eqnarray}
D=\left( \frac{c \rho}{E_\mathrm{mech}}\, \Omega_\mathrm{m} \int_{A_\mathrm{p}}\left|\Delta z \left(r\right)\right|^2dA\right)^{-1}
\label{eq:D}
\end{eqnarray}
is introduced, where $E_\mathrm{mech}$ is the total energy stored in the oscillator, $\rho$ denotes the density of silica, and $c$ the speed of sound in silica, $\Omega_\mathrm{m}/2 \pi$ is the mechanical oscillation frequency and $\Delta z \left(r\right)$ denotes the out of plane oscillation amplitude at position $r$, with the integration extending over the clamping area $A_\mathrm{p}$. $D$ 
\begin{figure*}
\includegraphics[scale=.9]{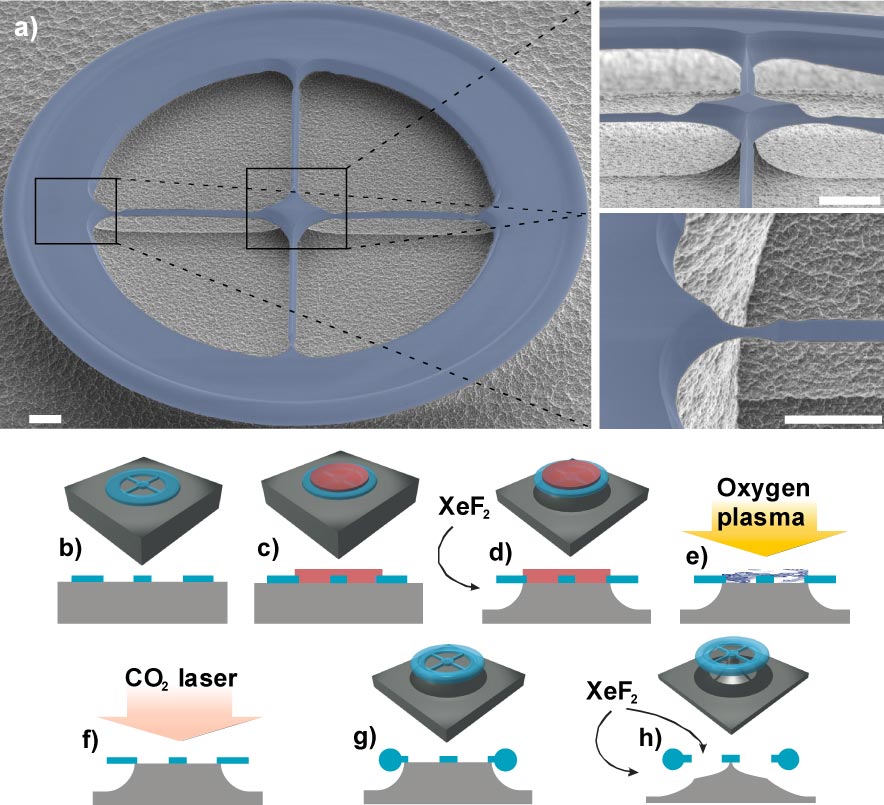}[t!]
\caption{\textbf{Novel optomechanical spoke-supported resonators. a} SEM images of on-chip optomechanical resonators consisting of a toroidal boundary supported by four spokes (scale bar: 5 $\mathrm{\mu m}$ in each panel). The optomechanical system (outer torus) is connected to the silicon chip via silica bridges, decoupling its radial motion from the clamping area in the center of the disk. For the depicted design mechanical $Q$-factors up to 80,000 are obtained at frequencies exceeding $30\,\mathrm{MHz}$.  \textbf{b-h} Fabrication process of ultra low dissipation optomechanical resonators on a chip. First the spokes pattern is defined in the oxide layer of an oxidized silicon chip by UV lithography and HF wet-etching (b). A protective layer of photoresist (c) covers the spokes while the structure is under-etched from the sides using an isotropic silicon dry etch (d). The photoresist is removed (e) and the laser reflow (f) creates the ultra-smooth toroid microcavity (g) while the spokes remain undamaged as their contact to the silicon pillar provides a sufficient heat sink. A final etching frees the spokes and creates the centered silicon support of the structure without affecting the ultra-high optical finesse (h).}
\label{f:3*}
\end{figure*}
is the mechanical quality factor that is expected if the clamping area $A_\mathrm{p}$ is modelled as a membrane, radiating acoustic energy with a power of $P=c \rho \, \Omega_\mathrm{m}^2 \int_{A_\mathrm{p}}\left|\Delta z\left(r\right)\right|^2 dA$. This acoustic loss then leads to a $Q$-factor of $D=\left(\frac{P}{\Omega_\mathrm{m} E_\mathrm{mech}}\right)^{-1}$ as given by equation (\ref{eq:D}). Note that a similar expression can be derived following a phonon tunneling approach\cite{Wilson-Rae08}.\\

\noindent
In Fig. 2 the $Q$-factors measured for two different modes of one microresonator are shown together with the simulated values of $D$. Strikingly, $D$ is directly proportional to the measured $Q$-factors, where $Q\approx 3 \cdot D$, revealing good agreement of measured mechanical loss and FEM simulations. This unambiguously shows that the mechanical quality factors are clamping loss limited. The 
\begin{figure*}[t!]
\includegraphics{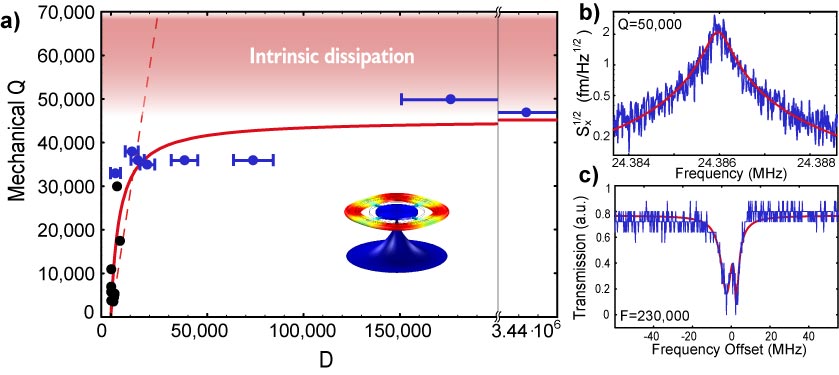}
\caption{\textbf{Ultra-low dissipation spokes-resonators. a} Measured mechanical $Q$-factors versus simulated $D$ values for the RBM of microresonators with spokes (blue dots) and different modes of high-$Q$ toroids (black dots) at room temperature. With the spokes resonators $Q$-factors of  50,000 can be achieved at frequencies above $20\, \mathrm{MHz}$. The data of the spoke-supported samples deviate from the linear behaviour demonstrated in Fig. \protect{\ref{f:2}} (indicated by dotted line). Instead they show a room temperature saturation (solid line) at $Q$-factors of $Q\approx 50,000$ which can be attributed to intrinsic losses as discussed in the text. The horizontal error bars mark the uncertainty in the calculation of $D$ due to deviations of the actual microresonator geometry from the geometry used in the simulation (inferred by SEM imaging). Inset: FEM simulation of the RBM of a spokes microresoantor. \textbf{b} Displacement spectrum of a spoke-supported resonator with a mechanical $Q$-factor of $Q=50,000$ at $24\, \mathrm{MHz}$. \textbf{c} Transmission spectrum of a spoke-supported microresonator showing optical mode-splitting \protect{\cite{KippenbergOL02}} with an unloaded finesse of 230,000.}
\label{f:4}\end{figure*}
prefactor of $3$ may be attributed to impedance matching conditions between the silica disk and the 
silicon pillar which are not taken into account in this model. Note that these, and all following measurements, were performed in a vacuum chamber ($p<1\, \mathrm{mbar}$) to fully eliminate the possible influence of gas damping (cf. Fig. 2). Besides explaining the origin of the mechanical dissipation of different modes and proving that plainly reducing the clamping area to a minimum is \textit{not} sufficient for optimized $Q$-factors the parameter $D$ moreover allows \textit{a-priori} modelling of the oscillator's mechanical quality. This constitutes a powerful, direct tool to design high-$Q$ mechanical oscillators with negligible clamping losses, which we have used to devise the novel structures described below.\\

\subsection{Ultra-low dissipation spoke-supported microresonators}
One such geometry is shown in Fig. 3a consisting of a toroid resonator with four spokes connecting it to the silicon support.
Incorporation of the silica spokes into a toroidal microresonator is possible by following the fabrication steps as shown in Fig. 3b-h. Importantly, the design allows independent control of the spokes' width, length and position for any major and minor diameter of the optical toroid microcavities yielding further degrees of freedom to tune their mechanical frequencies. Moreover, using the aforementioned FEM model of the clamping loss limited $Q$-factors, i.e. maximizing $D$, the geometry can be chosen such that the discrete eigenfrequencies of the spokes decouple the toroid's radial motion from the center of the silica disk, strongly mitigating clamping losses.\\
\noindent
Fig. \ref{f:4}a shows the dependence of the measured $Q$-factors on the simulated values of $D$ for several different samples of vastly different geometries with and without spokes. Indeed, maximizing $D$ using the spokes support allows obtaining reproducibly increased mechanical $Q$-factors. The quality factors which can be achieved at room temperature are boosted by approximately a factor of ten, e.g. a value of 50,000 is attained at $24\,\mathrm{MHz}$ as shown in Fig. \ref{f:4}b.
The corresponding  clamping loss limit as described by our model parameter $D$ can be pushed beyond the $10^6$-level using an optimized design. Yet, as shown in Fig. \ref{f:4}a a saturation of the $Q$-factors becomes evident and the data allow a fit according to a saturation model $Q^{-1}=1/(a D) + Q_\mathrm{sat}^{-1}$. Further loss mechanisms which will be addressed below become dominant for samples with strongly reduced clamping losses and--for room temperature operation--set an upper limit of $Q_\mathrm{sat} \approx 50,000$. The slope of the linear part $a$ obtained from this fit is slightly larger than the one obtained from Fig. \ref{f:2} (dashed line in Fig. \ref{f:4}a). This can be attributed to different impedance matching conditions as the sample of Fig. \ref{f:2} was 
fabricated using a $1 \,\mathrm{\mu m}$ silica preform whereas the new samples in Fig. \ref{f:4} consist of a $2 \,\mathrm{\mu m}$ silica layer.\\
\noindent
It is emphasized that the optical properties of the toroid microresonators are not affected by the 
optimized mechanical design. The $\mathrm{CO_2}$ laser assisted reflow process\cite{ArmaniNature03} creating the surface tension induced surface finish of the silica toroid can still be applied and the 
\begin{figure}[t!]
\includegraphics[scale=.85]{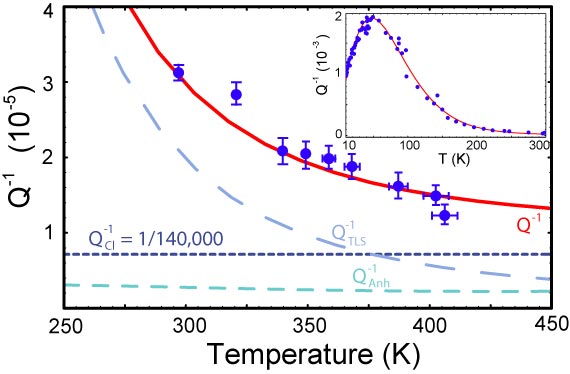}[t!]
\caption{\textbf{Temperature dependence of mechanical dissipation.} The damping $Q^{-1}$ of a $38\, \mathrm{MHz}$ sample as a function of temperature is shown. The damping decreases from $Q^{-1}=1/32,000$ at room temperature to a record value of $Q^{-1}=1/80,000$ at $410\, \mathrm{K}$. A fit (red line) including losses due to TLS ($Q_\mathrm{TLS}^{-1}$), anharmonicity ($Q_\mathrm{anh}^{-1}$) and a constant background ($Q_\mathrm{cl}^{-1}$) yields excellent agreement with the data and allows extracting an upper bound of $Q_\mathrm{cl}^{-1}=1/140,000$ for the remaining clamping losses. Further suppression of the intrinsic losses could enable $Q$-factors approaching the extracted clamping loss limit of $Q=140,000$. Inset: Damping $Q^{-1}$ of a different sample at $34\, \mathrm{MHz}$ illustrating the reliability of the fit (red line) of the three aforementioned contributions over a wide temperature range from $10\, \mathrm{K}$ to $300\, \mathrm{K}$.}
\label{f:5}
\end{figure}
subsequent $\mathrm{XeF_2}$ underetch does not affect the optical quality. As a result, the thermal bistability and optical mode-splitting\cite{KippenbergOL02}, salient features of ultra high-$Q$ silica microtoroids\cite{KippenbergAPL04}, can be observed in the novel spoke-supported microresonators. To illustrate this point, Fig. \ref{f:4}c shows the transmission spectrum of a spokes resonator with an unloaded optical finesse of $230,000$ displaying mode splitting\cite{KippenbergOL02}. At the same time, the mechanical design is not adversely affected by the laser reflow process, even for very small spokes (cf. Fig. 3). Thus, optical and mechanical properties of microtoroidal optomechanical resonators can be decoupled and optimized \textit{independently}. The mechanical frequencies and $Q$-factors are controlled by choosing appropriate spokes dimensions without affecting the choice of minor and major diameter of the toroid defining its optical frequencies and finesse\cite{KippenbergAPL04}.\\

\subsection{Temperature dependence}
In order to elucidate the apparent room temperature limit to the mechanical quality of $\approx 50,000$ and to quantify contributions of different origin, we studied the temperature dependence of the $Q$-factors. In Fig. \ref{f:5} the damping of the RBM is depicted as a function of temperature. The damping of $Q^{-1}=1/32,000$ measured for this sample at $38\,\mathrm{MHz}$ and room temperature continuously decreases when the temperature is increased. In particular, at $410\, \mathrm{K}$ a record value of $Q=80,000$ is attained, on par with the best published values in this frequency range\cite{Verbridge07}.
Besides demonstrating a further reduction of dissipation, the observed temperature dependence allows a detailed understanding of the limitations to the mechanical quality. As will be shown below, these are due to dissipation caused by two-level systems (TLS) common to amorphous materials\cite{Pohl02} and anharmonicity\cite{Vacher05}. The inset in Fig. \ref{f:5} shows the damping $Q^{-1}(T)$ obtained with a $34 \, \mathrm{MHz}$ sample measured from 10 to $300\, \mathrm{K}$\cite{Remiinprep}. A two parameter fit of the TLS losses together with the anharmonic contribution (no adjustable parameters) and a constant background (clamping losses) yields excellent agreement with the data over the full temperature range as well as with the parameter range reported in experiments conducted with bulk silica\cite{Vacher05,Rat05} (see methods). Using the same fit for the high temperature data ($295\, \mathrm{K}$ to $410\, \mathrm{K}$) displayed in Fig. \ref{f:5} also yields excellent agreement. Moreover, the extracted temperature independent background sets an upper bound for remaining clamping losses of $Q_\mathrm{cl}^{-1}=1/140,000$ experimentally confirming that a strong reduction of clamping losses is achieved. For this specific sample the $D$ model predicts six times lower clamping losses. Thus, it is likely that asymmetries in the structure cause additional clamping losses not taken into account in the above FEM model leading to the observed background of $1/140,000$. It is noted that thermoelastic damping\cite{ZenerPR37} for which an upper bound of $10^{-7}$ is found for the resonators considered here, can safely be neglected in this analysis. The intrinsic losses of silica (i.e. those caused by TLS and anharmonicity) are known to exhibit a local minimum at a temperature of around $500\, \mathrm{K}$\cite{Pohl02} and thus a further increase in temperature is expected to yield even higher $Q$-factors than those reported here.\\
The quantification of remaining clamping loss limit yielding values above $10^{5}$ is also very important for the prospects of future low temperature experiments, as the intrinsic losses in amorphous materials can be strongly reduced by cryogenic operation\cite{Pohl02}. After reaching a maximum at $\approx 50\,\mathrm{K}$ the intrinsic losses decrease again at lower temperatures (cf. data in Fig. \ref{f:4}). It is well known\cite{Pohl02} that the $Q$-factors reach a plateau of $Q\approx 3,000$ before rapidly increasing below a drop-off temperature $T_\mathrm{d}$ which shows a strong frequency dependence. For kHz frequencies, the drop-off temperature resides in the $\mathrm{mK}$-range\cite{Pohl02} and therefore silica is commonly not considered as a suitable material for low temperature oscillators. It is noted, however, that for high oscillation frequencies the drop-off temperature strongly increases and for e.g. $43\,\mathrm{MHz}$ ($T_\mathrm{d}=3\,\mathrm{K}$) in \cite{Pohl02} the intrinsic loss limited $Q$-factors were measured to reach 50,000 at $400\,\mathrm{mK}$ while quality factors exceeding 100,000 may be extrapolated for a temperature of $300\, \mathrm{mK}$\cite{Pohl02}. Thus, the spoke supported resonators are very promising optomechanical devices for cryogenic operation with the prospect of even improving the demonstrated outstanding room temperature $Q$-factors at low temperature, possibly approaching the extracted clamping loss limit beyond $Q=10^5$.\\

\section{Discussion}
\noindent
In summary, we presented novel micron-scale on-chip optomechanical resonators that for the first time allow independent control of optical and mechanical degrees of freedom within one and the same device. Avoided crossings of mechanical modes (as first predicted by Mindlin\cite{Mindlin51}) were observed which enabled a detailed understanding of mechanical dissipation, allowing the design of optomechanical systems combining mechanical $Q$-factors of 80,000, on par with the best published results in the frequency range above 30 MHz, with ultra-high optical finesse. This property yields improved performance for applications such as realizing narrow band filters and narrow line-width radiation pressure driven``photonic clocks''\cite{RokhsariIEEE}. Yet, it is also an essential ingredient in studies which seek to demonstrate quantum signatures of mesoscopic objects\cite{Caves81,SchwabPT05}. To illustrate this point, the ratio of displacement spectral density provided by radiation pressure quantum back-action and the thermal noise spectral density is given\cite{Tittonen99}: $S^\mathrm{ba}/S^\mathrm{th}\left[\Omega_\mathrm{m}\right] = \frac{4 \xi^2}{1 + 4 \Omega_\mathrm{m}^2/\kappa^2}\frac{\hbar\, Q\, \mathcal{F}^2}{\lambda (c/n)\, \pi\, k_B T\, m \Omega_\mathrm{m}}\,P$ ($\xi=1$ for a linear cavity and $\xi=\pi$ for a ring-cavity, $P$: power impinging on the cavity). For cryogenic operation at temperature $T=300\,\mathrm{mK}$, a $Q$-factor of 80,000 (which seems feasible given our results and the measurements at $400\, \mathrm{mK}$ in \cite{Pohl02}) and the realistic parameters $\lambda=1\, \mathrm{\mu m}$ (wavelength), $m=5\, \mathrm{ng}$ (effective mass), $\Omega_\mathrm{m}/2 \pi = 20 \, \mathrm{MHz}$ (mechanical frequency), $\mathcal{F}=300,000$ (finesse) and a cavity radius of $R=35 \, \mathrm{\mu m}$ (the cavity bandwidth $\kappa/2\pi$ for a toroidal cavity is then given by $\kappa=c/nR\mathcal{F}$, where $c$ denotes the speed of light in vacuo and $n=1.45$ the refractive index of silica) this expression reaches unity already at a launched power $P$ as low as $P=10\, \mathrm{\mu W}$. Thus, radiation pressure quantum back-action--which so far has remained experimentally inaccessible--may get within reach. Moreover, at a bath temperature of $T=300\,\mathrm{mK}$, and for a mechanical $Q$-factor of 80,000 already one tenth of the radiation-pressure cooling rates of $\Gamma_\mathrm{c}/2\,\pi = 1.56\, \mathrm{MHz}$ demonstrated in silica microtoroids\cite{SchliesserNP08} would enable laser cooling of the RBM in the resolved sideband regime to an occupation number of $n=\frac{k_B T/Q}{\hbar\Gamma_\mathrm{c}}\approx 0.5$, below unity, and would thus allow to venture into the regime of cavity \textit{quantum} optomechanics.\\
Pertaining to the wider implications of our work, the studied aspects of geometry dependent mechanical dissipation via mode coupling (which is also important for nanomechanical beam resonators\cite{SchwabP}) and the demonstrated quantification of clamping losses can be of broad interest and could directly be applicable to other whispering gallery mode resonators made of amorphous\cite{MaOL07,Eichenfield08} and crystalline material\cite{Grudinin06, Kiraz01} as well as to a wide range of other micro- and nanomechanical oscillators, such as photonic crystal based systems\cite{Notomi06} which are also expected to exhibit optomechanical coupling.

\section{Methods}
\subsection{Optomechanical monitoring}
To monitor the mechanical modes' thermal noise, laser light is coupled to the high-$Q$ optical whispering gallery mode of the silica resonators using a tapered optical fibre. An external cavity diode laser is locked detuned by approximately half the optical line-width $\kappa$ with respect to a cavity resonance $\omega_\mathrm{0}$. Thus, the small, thermal fluctuations $\Delta x$ of the cavity radius cause changes in the optical path-length which imprint themselves linearly into an amplitude modulation $\Delta P$ of the power transmitted by the cavity, i.e. $\Delta P \propto \Delta x$. Spectral analysis of the transmitted power thus allows extracting the mechanical frequencies $\Omega_\mathrm{m}/2\pi$ and dissipation rates $\Gamma_\mathrm{m}/2 \pi$ from which the corresponding quality factor $Q=\Omega_\mathrm{m}/\Gamma_\mathrm{m}$ is inferred. To minimize radiation pressure induced back-action, i.e. to rule out any amplification\cite{KippenbergPRL05} or cooling\cite{SchliesserPRL06} which would lead to a modification of the natural line-width\cite{SchliesserPRL06}, the laser power used in these measurements is kept at a very low level, well below the threshold for the parametric instability\cite{KippenbergPRL05}. As an additional check the measurements are always performed both on the red and blue detuned wing of the optical resonance. The $Q$-factors from both measurements usually agree within less than $5\%$, indicating negligible radiation pressure backaction.\\
\noindent
In order to achieve a higher displacement sensitivity allowing to observe further mechanical modes of the microresonators, and in particular the avoided crossing in Fig. \ref{f:2}b), an adapted version of the H\"ansch-Couillaud polarization spectroscopy\cite{HaenschOC80} using a Nd:YAG laser locked to cavity resonance is employed\cite{NJPsubmission}. The polarization of the field injected into the taper is chosen such that only a small portion couples to a polarization non-degenerate whispering gallery mode of the cavity while the larger part bypasses the cavity serving as a local oscillator. Analyzing the phase-shift of the field exiting the cavity via a homodyne detection a shot-noise limited sensitivity $\sqrt{S_x^{sn}\left[\Omega_\mathrm{m}\right]}=\frac{\lambda}{8 \pi \mathcal{F}}\sqrt{\frac{\hbar \omega}{P}}\sqrt{1 + 4\Omega_\mathrm{m}^2/\kappa^2}$ of order $10^{-19}\, \mathrm{m}/\sqrt{\mathrm{Hz}}$ is achieved at the frequencies of interest from 10-100 MHz\cite{NJPsubmission}, where $P$ denotes the power impinging on the cavity. Thus, not only the radial breathing mode but more than twenty different mechanical modes of the toroid between 0-$100\,\mathrm{MHz}$ can be monitored. All observed modes can be identified with the according mode pattern obtained by FEM simulation with a relative frequency deviation of less than $5\%$\cite{NJPsubmission}. 
\subsection{FEM simulations}
A commercial software package (COMSOL Multiphysics) is used for 3D finite element modelling of the cavity structures. Using fixed boundary conditions at the bottom of the conical silicon pillar, the eigenvalue solver provided by the package calculates the eigenfrequencies $\Omega_\mathrm{m}/2 \pi$ and the corresponding stress and strain fields of the meshed structure by diagonalizing its coupled linear equations of motion. Summing over all elements the total mechanical energy $E_\mathrm{mech}$ of the mode can be calculated. Typically 100,000 elements are used where in the clamping region the characteristic size of an element is of order $200\,\mathrm{nm}$. In order to calculate the parameter $D$, the squared out of plane displacement amplitude $\left|\Delta z\left(r\right)\right|^2$ is summed over the plane connecting the silica resonator and the silicon substrate.\\
\subsection{Intrinsic losses in silica}
For fitting the data in Fig. 5 we employed the known model for losses caused by two level systems $Q^{-1}_\mathrm{TLS}$ and anharmonic damping $Q^{-1}_\mathrm{anh}$. The expression for $Q^{-1}_\mathrm{TLS}$ originating from microscopic two level systems described by a distribution of double-well potentials reads\cite{Vacher05}:
\begin{eqnarray}
Q^{-1}_\mathrm{TLS}&=&C\cdot \mathrm{Erf}\left(\frac{\sqrt{2} T}{\Delta_\mathrm{c}}\right)\frac{1}{T} \nonumber \\
&\hphantom{=}& \int_0^{\infty}\left(\frac{V}{V_\mathrm{0}}\right)^{-\zeta}e^{-\frac{V^2}{2 V_\mathrm{0}^2}}\frac{\Omega \tau_\mathrm{0} e^{V/T}}{1+\Omega^2 \tau_\mathrm{0}^2 e^{2V/T}}\,dV\,,
\end{eqnarray}
where $\mathrm{Erf}(z)=\frac{2}{\sqrt{\pi}}\int_0^{z}e^{-x^2}\,dx$ is the error function, $T$ denotes the temperature and $\Omega/2\pi$ the oscillation frequency. $V_\mathrm{0}$ is a scale for the distribution of the potential barrier $V$ separating both wells, and $\Delta_\mathrm{c}$ scales the asymmetry in the depth of the wells (all in temperature units). $\zeta$ is a dimensionless exponent, $\tau_\mathrm{0}$ a relaxation time and $C$ a dimensionless parameter. Measurements spanning from $11\,\mathrm{kHz}$ to above $200\,\mathrm{MHz}$ as well as from a few up to a few hundred Kelvin allow an excellent fit using the parameters $\zeta=0.28\pm 0.03$, $V_\mathrm{0}=667\pm21\,\mathrm{K}$, $V_\mathrm{0}/\Delta_\mathrm{c}=7.7\pm0.7$, $\mathrm{log}_{10}\,\tau_\mathrm{0}=-12.2\pm0.18$ and $C=\left(1.45 \pm0.35\right)\cdot 10^{-3}$\cite{Vacher05}. For the fit presented in this work we adopt the above mean values for $\zeta$, $V_\mathrm{0}$ and $V_\mathrm{0}/\Delta_\mathrm{c}$ while using $\tau_\mathrm{0}$ and $C$ as  fit parameters. The fitted values are $\mathrm{log}_{10}\,\tau_\mathrm{0}=-12.1$ (low temperature fit at $34\,\mathrm{MHz}$), $\mathrm{log}_{10}\,\tau_\mathrm{0}=-12.05$ (high temperature fit at $38\,\mathrm{MHz}$) and $C=1.8\cdot 10^{-3}$, falling within the parameter range as specified in\cite{Vacher05}.
The anharmonic contribution $Q_\mathrm{anh}^{-1}$ is given by\cite{Vacher05}:
\begin{eqnarray}
Q_\mathrm{anh}^{-1}=\gamma^2\,\frac{C_\mathrm{v}(T)\, v(T)\, T}{2 \rho \, v_\mathrm{D}^3\,(T)} \frac{\Omega \, \tau_\mathrm{th}(T)}{1 + \Omega^2 \tau_\mathrm{th}^2(T)}.
\end{eqnarray}
Here, $v_\mathrm{D}$ denotes the Debye and $v$ the sound velocity. For the fit in Fig. 5 we adopt $v_\mathrm{D}^3=0.322\, v^3$\cite{Rat05}, the Gr\"uneisen parameter $\gamma^2=3.6$\cite{Vacher05}, the mean lifetime of thermal phonons $\tau_\mathrm{th}(T)$ determined in\cite{Vacher05} and the known values for the specific heat per unit volume $C_\mathrm{v}(T)$, the sound velocity $v(T)$ and the density $\rho$ of silica (e.g. \cite{Bansal86}) without any fit parameter.\\

\acknowledgments
\noindent
TJK acknowledges support via an Independent Max Planck Junior Research Group Grant, a Marie Curie Excellence Grant (JRG-UHQ), the DFG funded Nanosystems Initiative Munich (NIM) and a Marie Curie Reintegration Grant (RG-UHQ). The authors gratefully acknowledge J. Kotthaus, A. Rogach and F. B\"ursgens for access to clean-room facilities for micro-fabrication, A. Marx for assistance in scanning electron microscopy and I. Wilson-Rae for stimulating discussions.

\end{document}